\documentclass[a4paper]{PoS}

\title{The linear velocity field of SDSS DR7 galaxies: constraints on flow amplitudes and the growth rate}

\ShortTitle{The linear velocity field of SDSS DR7 galaxies}

\author{\speaker{Martin Feix}\\
        Department of Physics, Israel Institute of Technology - Technion, Haifa 32000, Israel\\
        E-mail: \email{mfeix@physics.technion.ac.il}}

\abstract{Large-scale peculiar motion modulates the observed luminosity distribution of galaxies. Using about half a million SDSS
galaxies, this can be harnessed to obtain bounds on peculiar velocity moments, the amplitude of the linear matter power spectrum,
$\sigma_{8}$, and the growth rate of density perturbations at $z\sim 0.1$. Results obtained from this approach agree well with the
predictions of the $\Lambda$CDM model and are consistent with the reported $\sim 1$\% zero-point tilt in the SDSS photometry.}

\FullConference{Frontiers of Fundamental Physics 14 - FFP14,\\
                15-18 July 2014\\
                Aix Marseille University (AMU) Saint-Charles Campus, Marseille }

\begin{document}
\section{Introduction}
If the standard paradigm of cosmology is correct, the observed structure of the universe originated from tiny density
fluctuations via gravitational instability. The clustering process is inevitably associated with peculiar motions of matter, i.e.
deviations from a pure Hubble flow, which exhibit a coherent pattern on large scales. Since galaxies can, to good approximation,
be treated as test particles, they should appropriately reflect the underlying peculiar velocity field which contains valuable
information for constraining and discriminating between different cosmological models.

Redshifts of galaxies are systematically altered by the line-of-sight components of their peculiar velocities and differ from
their actual distances. Consequently, intrinsic galaxy luminosities inferred from the observed flux using redshifts rather than
distances appear brighter or dimmer. Since this effect is obscured by the natural scatter in the distribution of magnitudes, it
cannot be used to derive the peculiar velocities of individual galaxies. However, it is possible to approach peculiar velocities
in a statistical sense by constraining the parameters of some appropriate, predefined velocity model. For instance, constraints
on the bulk motion of galaxies in a subvolume of a given survey can be derived by comparing the luminosity distribution of
galaxies in the subvolume with that of the whole survey. Reconstructing the linear velocity field from the observed density field
in redshift space, this technique further provides a way of estimating the growth rate of density perturbations which is independent
from the apparent clustering anisotropy of galaxies \cite{Kaiser1987, Hamilton1998}.

Below I briefly review the luminosity-based approach outlined above and discuss its recent application to galaxy data from the
Sloan Digital Sky Survey (SDSS) \cite{York2000}. Methods of this kind have first been adopted to estimate the motion of Virgo
relative to the Local Group \cite{Tammann1979}, and more recently, to constrain bulk flows and the growth rate in the local
universe within $z\sim 0.01$ \cite{Nusser2011, Nusser2012}.

\section{Velocity-induced modulation of observed galaxy luminosities}
Due to inhomogeneities, the observed redshift $z$ of a galaxy is generally different from its cosmological one, $z_{c}$, which
is defined for the homogeneous background. In linear perturbation theory, the two quantities are connected through \cite{SW1967}
\begin{displaymath}
\frac{z-z_{c}}{1+z} = \frac{V(t,r)}{c} - \frac{\Phi(t,r)}{c^2} - \frac{2}{c^2}\int_{t(r)}^{t_{0}}{\rm d}t
\frac{\partial\Phi\left\lbrack\hat{\textbf{\textit{r}}}r(t),t\right\rbrack}{\partial t}\approx \frac{V(t,r)}{c},
\end{displaymath}
where $\hat{\textbf{\textit{r}}}$ denotes a unit vector along the line of sight to the galaxy, $V$ is the galaxy's (physical) radial
peculiar velocity, and $\Phi$ the usual gravitational potential. Focusing on low redshifts at the present time, the velocity $V$ is
explicitly assumed as the dominant contribution. All fields are considered relative to their present-day values at $t_{0}$. The
difference in redshifts enters the distance modulus ${\rm DM}=25+5\log_{\rm 10}\lbrack D_{L}/{\rm Mpc}\rbrack$ and causes the observed
absolute magnitudes $M$ to deviate from their true values $M^{(t)}$, i.e.
\begin{displaymath}
M = m - {\rm DM}(z) - K(z) + Q(z) = M^{(t)} + 5\log_{10}\frac{D_{L}(z_{c})}{D_{L}(z)},
\end{displaymath}
where $m$ is the apparent magnitude, $D_{L}$ is the luminosity distance, and the functions $Q(z)$ and $K(z)$ account for luminosity
evolution and $K$-correction \cite{Blanton2007}, respectively. Restricted to scales where linear theory is valid, $M-M^{(t)}$ varies
systematically over the sky, and thus provides a probe of the cosmic velocity field.

Given a galaxy survey with magnitudes, (spectroscopic) redshifts, and angular positions $\hat{\textbf{\textit{r}}}_{i}$ on the sky,
the idea is to choose an appropriate parameterized model of $V(\hat{\textbf{\textit{r}}}, z)$ and to maximize the probability of the
data,
\begin{displaymath}
P_{\rm tot} = \prod\limits_{i}P\left (M_{i}\vert z_{i}, V(\hat{\textbf{\textit{r}}}_{i},z_{i})\right ) =
\prod\limits_{i}\left (\phi(M_{i})\middle /\int_{M_{i}^{+}}^{M_{i}^{-}}\phi(M){\rm d}M\right ),
\end{displaymath}
where redshift errors are neglected \cite{Nusser2011}. Here $\phi(M)$ denotes the galaxy luminosity function (LF), and the limiting
magnitudes $M^{\pm}$ depend on the cosmological redshift $z_{c}$, and hence on $V(\hat{\textbf{\textit{r}}}, z)$. The motivation of
this approach is to find those velocity model parameters which minimize the spread in the observed magnitudes.

\section{Constraints from SDSS galaxy luminosities at z {\textasciitilde} 0.1}
Galaxy data from the SDSS Data Release 7 \cite{Abazajian2009} trace the cosmic velocity field out to
$z\sim 0.1$. Here I summarize recent results obtained from applying the luminosity method to suitable subsets comprising up to half
a million galaxies \cite{Feix2014, Feix2015}.

{\underline{\it Data}}.---The analysis is based on the latest version of the NYU Value-Added Galaxy Catalog (NYU-VAGC), adopting the
subsample {\tt safe} to minimize incompleteness and systematics \cite{Blanton2005}. Using Petrosian $^{0.1}r$-band magnitudes, only
galaxies with $14.5 < m_{r} < 17.6$, $-22.5 < M_{r} - 5\log_{10}h < -17.0$, and $0.02 < z < 0.22$ (relative to the CMB frame) are
included. In addition, galaxy mock catalogs were built to study errors and known systematics of the data.

{\underline{\it Radial velocity model}}.---Constraints on velocity moments and derived quantities assume a bin-averaged model
$\tilde{V}(\hat{\textbf{\textit{r}}})$ in two redshift bins, $0.02 < z < 0.07$ and $0.07 < z < 0.22$. For each bin, the velocity
field was decomposed into spherical harmonics, i.e.
\begin{displaymath}
a_{lm} = \int{\rm d}\Omega\tilde{V}(\hat{\textbf{\textit{r}}})Y_{lm}(\hat{\textbf{\textit{r}}}),\qquad \tilde{V}(\hat{\textbf{\textit{r}}})
= \sum\limits_{l,m}a_{lm}Y_{lm}^{*}(\hat{\textbf{\textit{r}}}),\qquad l>0,
\end{displaymath}
where the sum over $l$ is cut at some maximum value $l_{\rm max}$. Because the SDSS data cover only part of the sky, the inferred $a_{lm}$
are not statistically independent. The impact of the angular mask was studied with the help of suitable galaxy mock catalogs. The monopole
term ($l=0$) was not included since it is degenerate with an overall shift of magnitudes.

{\underline{\it LF estimators}}.---Reliable measurements of the galaxy LF form a key step in the analysis. To assess the robustness of
results with respect to different LF models, the data were examined using LF estimators based on a Schechter form and a more flexible
spline-based model, together with several combinations and variations thereof. For simplicity, a linear dependence of the luminosity
evolution with redshift was assumed.

{\underline{\it Bulk flows and higher-order velocity moments}}.---Accounting for known systematic errors in the SDSS photometry, measurements
of ``bulk flows'' are consistent with a standard $\Lambda$CDM cosmology at a $1$--$2\sigma$ confidence level in both redshift bins. A joint
analysis of the corresponding three Cartesian components confirms this result. To characterize higher-order moments, direct constraints on
the angular velocity power spectrum $C_{l} = \langle |a_{lm}|^{2}\rangle$ were obtained up to the octupole contribution. The estimated
$C_{l}$ were found compatible with the theoretical power spectra of the $\Lambda$CDM cosmology.

\begin{figure}
\hspace*{0.12\linewidth}
\includegraphics[width=0.8\linewidth]{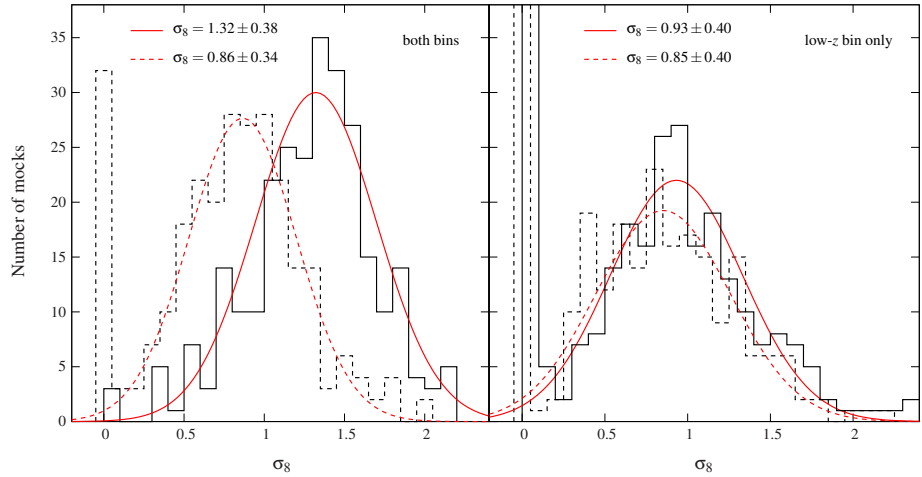} 
\caption{Estimates of $\sigma_{8}$ from galaxy mocks: shown are the recovered distributions and respective Gaussian fits with (solid
lines) and without (dashed lines) the inclusion of a (randomly oriented) photometric tilt, using the information in both redshift bins
(left) and the low-$z$ bin only (right).}
\label{fig1}
\end{figure}

{\underline{\it Constraints on $\sigma_{8}$}}.---Assuming a prior on the $C_{l}$ as dictated by the $\Lambda$CDM model with fixed
Hubble constant and density parameters, the amplitude of the velocity field was estimated in terms of $\sigma_{8}$. Because of a
dipole-like tilt in the galaxy magnitudes \cite{Padmanabhan2008}, raw estimates of $\sigma_{8}$ were biased toward larger values
(see Figure \ref{fig1}). After correcting for this magnitude tilt, it was found that $\sigma_{8} = 1.1\pm 0.4$ for the combination
of both redshift bins and $\sigma_{8} = 1.0\pm 0.5$ for the low-$z$ bin only, where the low accuracy is due to the limited number
of galaxies.

\begin{figure}
\hspace*{0.02\linewidth}
\includegraphics[width=0.5\linewidth]{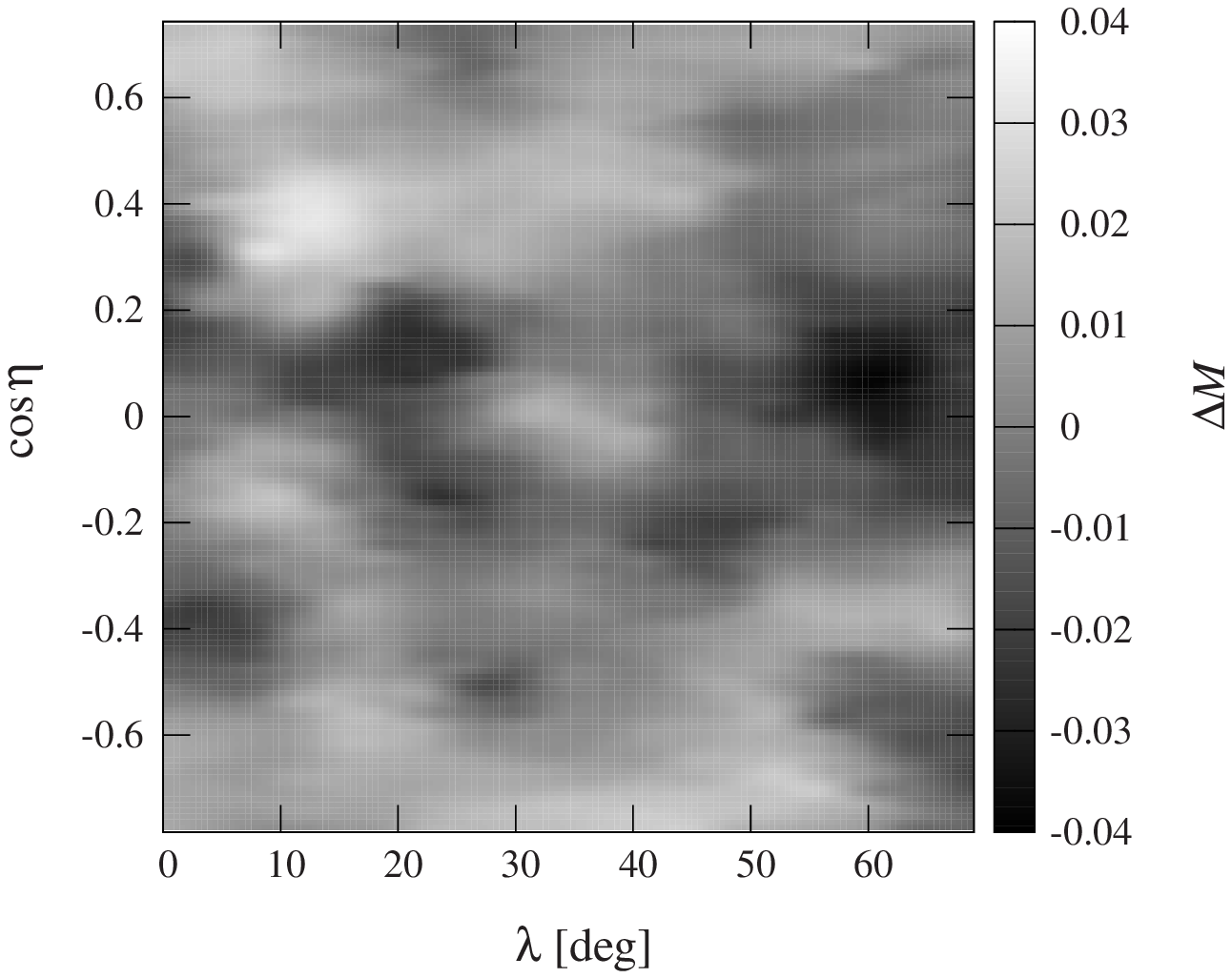}
\hspace*{0.09\linewidth}
\includegraphics[width=0.41\linewidth]{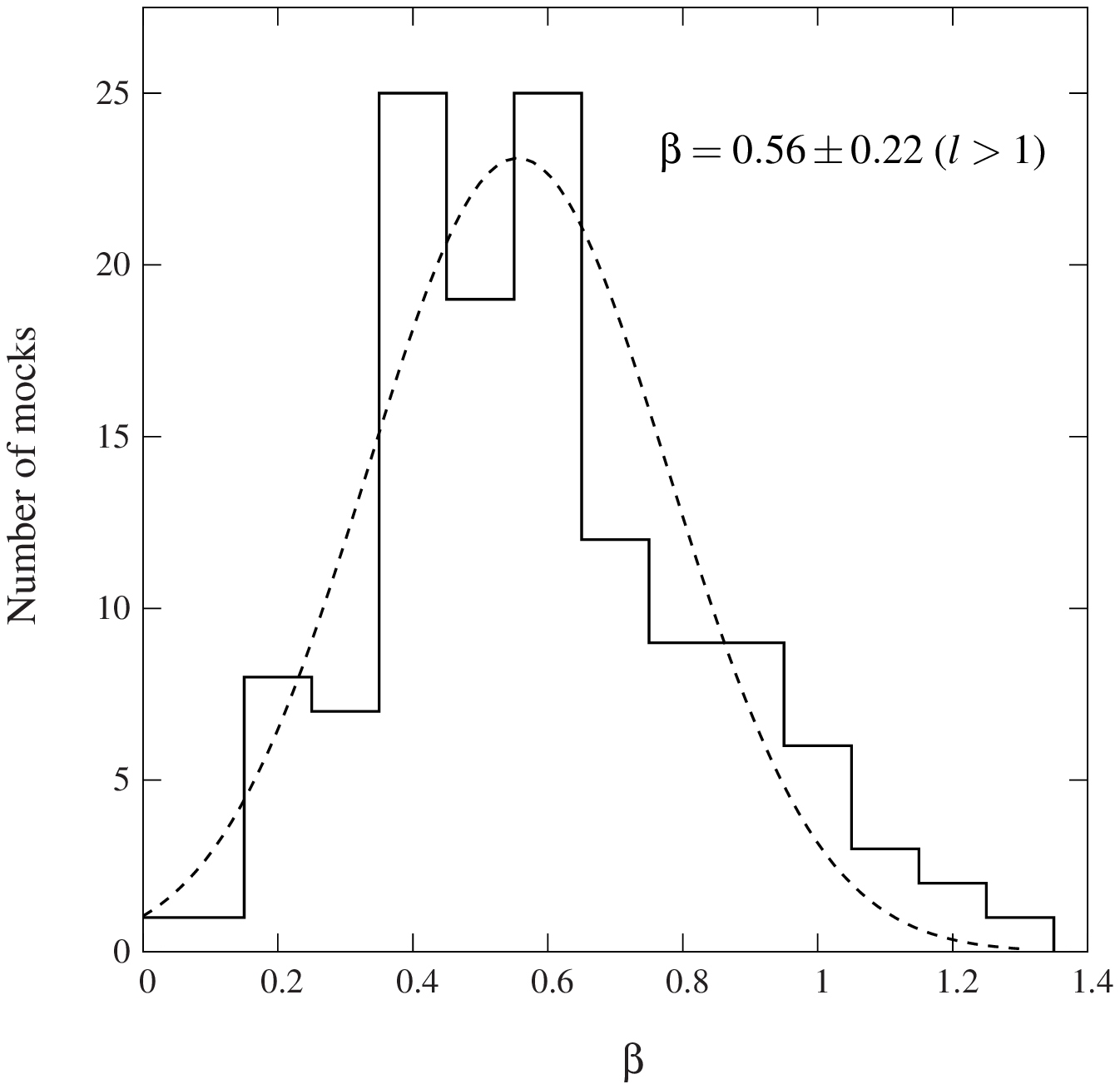}
\caption{{\underline{\textit{Left panel.}}}---Velocity-induced modulation of galaxy magnitudes at $z=0.1$ over an angular patch expressed
in SDSS survey coordinates (based on a random mock). {\underline{\textit{Right panel.}}}---Distribution of $\beta$-estimates (and Gaussian
fit) derived from mock catalogs, excluding the dipole in the velocity reconstruction ($l>1$).}
\label{fig2}
\end{figure}

{\underline{\it The linear growth rate}}.---Modeling the linear velocity field from the observed galaxy clustering in redshift space,
the luminosity approach is capable of constraining the growth rate of density perturbations, $\beta = f(\Omega)/b$, where $b$ is the
linear galaxy bias \cite{Nusser2012}. To this end, both magnitude- and volume-limited subsamples were selected over a rectangular
patch (in survey coordinates) within the range $0.06 < z < 0.12$. Following \cite{Nusser1994}, the velocity reconstruction was done
in ``harmonic'' space by smoothing the density field with a Gaussian kernel of $10h^{-1}$ Mpc radius and solving
\begin{displaymath}
\frac{1}{s^{2}}\frac{\rm d}{{\rm d}s}\left (s^{2}\frac{{\rm d}\Phi_{lm}}{{\rm d}s}\right )
- \frac{1}{1+\beta}\frac{l(l+1)\Phi_{lm}}{s^{2}} = \frac{\beta}{1+\beta}\left (\delta^{g}_{lm}
- \frac{{\rm d}\log S}{{\rm d}s}\frac{{\rm d}\Phi_{lm}}{{\rm d}s}\right ),
\end{displaymath}
where $S$ is the selection function and $\Phi(s)$ is the velocity potential expressed in redshift space. Boundary conditions were fixed
by setting the density contrast outside the observed volume to zero. An example of how the full velocity field modulates galaxy magnitudes
at $z=0.1$ is depicted in Figure \ref{fig2} (left panel). The velocity reconstructions and the method's accuracy were assessed with the help
of mocks generated from the Millennium Simulation \cite{Springel2005, Henriques2012}. Excluding the dipole in the velocity reconstruction
($l>1$), the found distribution of $\beta$-estimates peaks at the true value $\beta_{\rm true}=0.52$, deviating from a symmetric Gaussian
mainly because of the velocity field's nonlinear dependence on $\beta$ (see right panel of Figure \ref{fig2}). A preliminary analysis of the
SDSS data with fixed Hubble constant and density parameters from \cite{Calabrese2013} yields $\beta=0.42\pm 0.14$ which, using the power
spectrum amplitude of $L^{\star}$-galaxies from \cite{Tegmark2004}, translates into $f\sigma_{8}=0.37\pm 0.13$ and is consistent with the
$\Lambda$CDM model. For samples with similar characteristics, the accuracy is comparable to what is obtained from recent measurements of
anisotropic clustering in redshift space \cite{Howlett2014}.

\section{Outlook}
Current and next-generation spectroscopic galaxy surveys are designed to reduce data-inherent systematics because of larger sky coverage
and improved photometric calibration in ground- and space-based experiments \cite{Levi2013, Laureijs2011}. Together with the above results,
these observational perspectives give confidence that the luminosity-based approach will be established as a standard cosmological probe,
independent from and complementary to the more traditional ones based on galaxy clustering, gravitational lensing and redshift-space
distortions. The method considered here does not require accurate redshifts and can also be used with photometric redshift surveys such as the
2MASS Photometric Redshift catalog (2MPZ) \cite{Bilicki2014} to recover signals on scales larger than the spread of the redshift error.

\end{document}